\documentclass[10pt,twocolumn,letterpaper]{article}

\usepackage{cvpr}
\usepackage{times}
\usepackage{epsfig}
\usepackage{graphicx}
\usepackage{amsmath}
\usepackage{amssymb}
\usepackage{url}
\usepackage{stfloats}
\usepackage{multirow}   

\usepackage[pagebackref=true,breaklinks=true,letterpaper=true,colorlinks,bookmarks=false]{hyperref}

 \cvprfinalcopy 

\ifcvprfinal\fi

\begin{document}
\title{End-to-end Optimized Video Compression with MV-Residual Prediction}

\author{XiangJi Wu$^1$, Ziwen Zhang$^1$, Jie Feng$^1$, Lei Zhou$^1$, Junmin Wu$^1$\\
$^1$Tucodec Inc\\
{\tt\small \{wuxiangji\}@tucodec.com}
}


\maketitle

\begin{abstract}
  We present an end-to-end trainable framework for P-frame compression in this paper. A joint motion vector (MV) and residual prediction network MV-Residual is designed to extract the ensembled features of motion representations and residual information by treating the two successive frames as inputs. The prior probability of the latent representations is modeled by a hyperprior autoencoder and trained jointly with the MV-Residual network. Specially, the spatially-displaced convolution is applied for video frame prediction, in which a motion kernel for each pixel is learned to generate predicted pixel by applying the kernel at a displaced location in the source image. Finally, novel rate allocation and post-processing strategies are used to produce the final compressed bits, considering the bits constraint of the challenge. The experimental results on validation set show that the proposed optimized framework can generate the highest MS-SSIM for P-frame compression competition.\\
\end{abstract}

\section{Introduction}
Recently, artificial neural networks (ANNs) have been applied to solve the image and video compression problem and a number of works have been proposed \cite{minnen2018joint,lee2018context,chen2019learning,cheng2019learning,habibian2019video,lu2019dvc,yang2020learning}. 
Recent studies in deep learning based image compression methods have achieved significant performance improvement and they focus on designing end-to-end optimized frameworks \cite{balle2018variational,theis2017lossy,minnen2018joint,lee2018context}, in which the modules such as transformation, quantization and entropy estimation are optimized jointly. It is therefore not surprising to see that Deep Neural Networks (DNN) have attracted attention for solving video compression tasks. Many approaches \cite{xu2018reducing,liu2018one,li2019densenet} were proposed to replace the components in traditional video codecs by DNNs. For example, Liu \emph{et al} \cite{liu2018one} utilized a DNN in the fractional interpolation of motion compensation, and \cite{li2019densenet} applied DNNs to improve performance of the in-loop filter. End-to-end video compression frameworks have also been studied widely and various approaches have been proposed. For example Wu \emph{et al.} presented a framework for predicting frames by interpolation from reference frames, and then the image compression network of was applied to compress the residual. In 2019, Lu \emph{et al.} \cite{lu2019dvc} proposed the Deep Video Compression (DVC) method, in which optical flow was used to predict the temporal motion, and then two compression subnetworks were designed to compress the motion and residual. In order to realize spatial-temporal energy compaction in learning image and video compression,  a spatial-temporal energy compaction was incorporated into the loss function to improve the video compression performance in \cite{cheng2019learning}. Meanwhile,  Habibian \emph{et al.} \cite{habibian2019video} firstly employed a model that consisted of a 3D autoencoder with a discrete latent space and an autoregressive prior for video compression. In \cite{chen2019learning}, the concept of Pixel-MotionCNN (PMCNN) which includes motion extension and hybrid prediction networks was proposed to design more robust motion prediction moduel. PMCNN can model spatiotemporal coherence to effectively perform predictive coding inside the learning network. Different from these methods which are trained with one loss function applied on all frames, Yang \emph{et al.} \cite{yang2020learning} proposed a Hierarchical Learned Video Compression (HLVC) method with three hierarchical quality layers and a recurrent enhancement network. The video frames are compressed in the hierarchical layers 1, 2 and 3 with decreasing quality, using an image compression method for the first layer and the proposed BDDC and SMDC networks for the second and third layers, respectively.\\
\indent In CLIC 2020 P-frame compression challenge, we propose a novel video compression framework which consists of a MV-Residual prediction network for video framework prediction and a post-processing module for visual quality enhancement. The MV-Residual prediction network is capable of estimating motion vectors and residual information simultaneously. Moreover, the techniques such as hyperprior base rate estimation, soft quantization and resource allocation which were proposed by Balle et al. and Mentzer et al. \cite{balle2018variational,mentzer2018conditional} have also been utilized to improve the compression performance. 

 \begin{figure*}
   \begin{center}
   \centerline{\includegraphics[width=16cm,height=6.3cm]{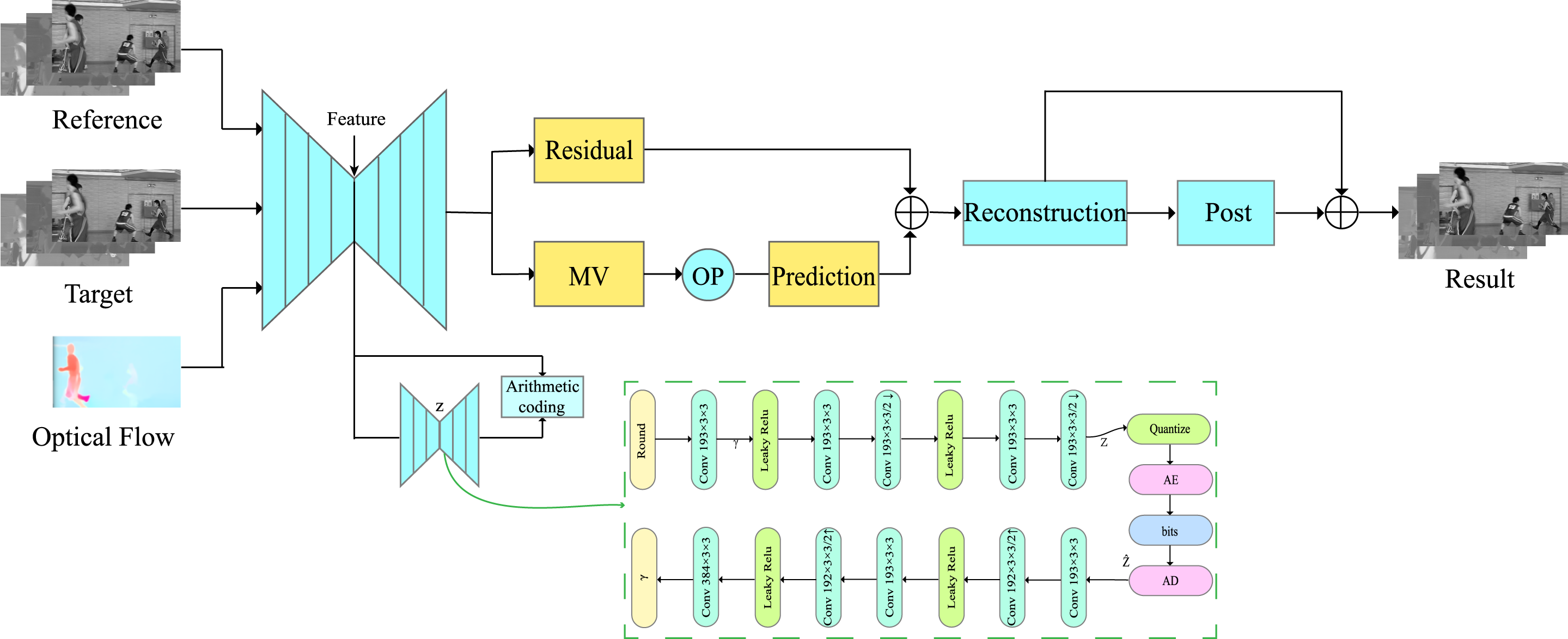}} %
 \end{center}
  \caption{Illustration of the proposed architecture. Input data is fed into the MV Encoder-decoder network and the hyperprior network works after encoding. Details for hyperprior autoencoder is shown in the dotted frame, convolution parameters are denoted as number of filter $\times$ kernel height $\times$ kernel width / down or upsampling stride, where $\downarrow$ indicates downsampling and $\uparrow$ indicates upsampling. AE, AD represent arithmetic encoder and arithmetic decoder. OP operation stands for SDC-Net. \cite{reda2018sdcnet}}
 \label{fig:aec}
 \end{figure*}
\section{End-to-end Optimized Video Compression with MV-Residual Prediction
}
\subsection{Overview of the Proposed Model}
\indent The proposed model is based on our CVPR 2019 CLIC framework in Low-rate compression \cite{zhou2019end}.  Fig \ref{fig:aec} provides an overview of our end-to-end video compression architecture, which can be optimized in an end-to-end manner. The brief summarization on the working process is introduced as follows:\\
\indent \textbf{Step1. Data preprocessing.} In this year's competition, the validation and test sets are subsets of the training set. Thus, any unnecessary modifications to the training data would not benefit the final result. To gain better performance and get rid of any unnecessary loss, we simply up-sampled the given pictures with format of YUV420 to YUV444 as reference frame and target frame. The optical flow is generated with pretrained PWC-Net  \cite{reda2018sdcnet}.\\
\indent \textbf{Step2. MV Encoder-decoder network.} In order to encode the motion information, we design an auto-encoder style CNN for better encoding. In this step, after a series of convolution operations and nonlinear transformations, representations of the motion will be generated. Then, the latent feature is quantized and fed into a hyperprior autoencoder to obtain prior probability, which is fully discussed in Section \ref{REM}.\\
\indent \textbf{Step3. Motion estimation.} An OP operation is designed to obtain the prediction based on the motion vector calculated by the previous network. To avoid blurry reconstruction, we utilized SDC-net \cite{reda2018sdcnet} as the OP operation, which is fully differentiable and thus allows our model to train end-to-end. More information is provided in Section \ref{SDC}.\\
\indent \textbf{Step4. Post processing.} With reconstruction obtained by summing residual and prediction, we employ a ResNet style network to optimize our result to get final output.
\indent \textbf{Step5. Variable rate.} In order to make full use of every bit space available, a rate control module \cite{choi2019variable} is utilized to fit our model to variable compression rates with a single set of weights. A rate control parameter, Lagrange multiplier is used as a conditional input for our end-to-end model, and contributes to our loss function for better optimization.

\subsection{Spatially-displaced Convolution}\label{SDC}
Given the reference frame and decoded motion vector produced by previous stage, a simple way to estimate motion is to calculate vector-based transformation. However, as discussed in \cite{reda2018sdcnet}, such operation will lead to insufficient representation of the motion and end in blurry results. Following \cite{reda2018sdcnet}, we use SDC-net to estimate motion:
\begin{equation}
I_{t+1} = \mathcal T( \mathcal G (I_{1:t},F_{2:t}),I_t),
\end{equation}
where $\mathcal T$  is realized with SDC operating on the reference frame $I_t$ and $F_i$ refers to decoded optical flow. $\mathcal G$ is a fully convolutional network which takes in a sequence of past frames $I_{1:t}$ and outputs pixel-wise separable kernels ${K_u,K_v}$ In this way, predictions of multiple frames will be extended naturally by recirculating new inputs.
 \begin{figure}
   \begin{center}
   \centerline{\includegraphics[width=7cm,height=4cm]{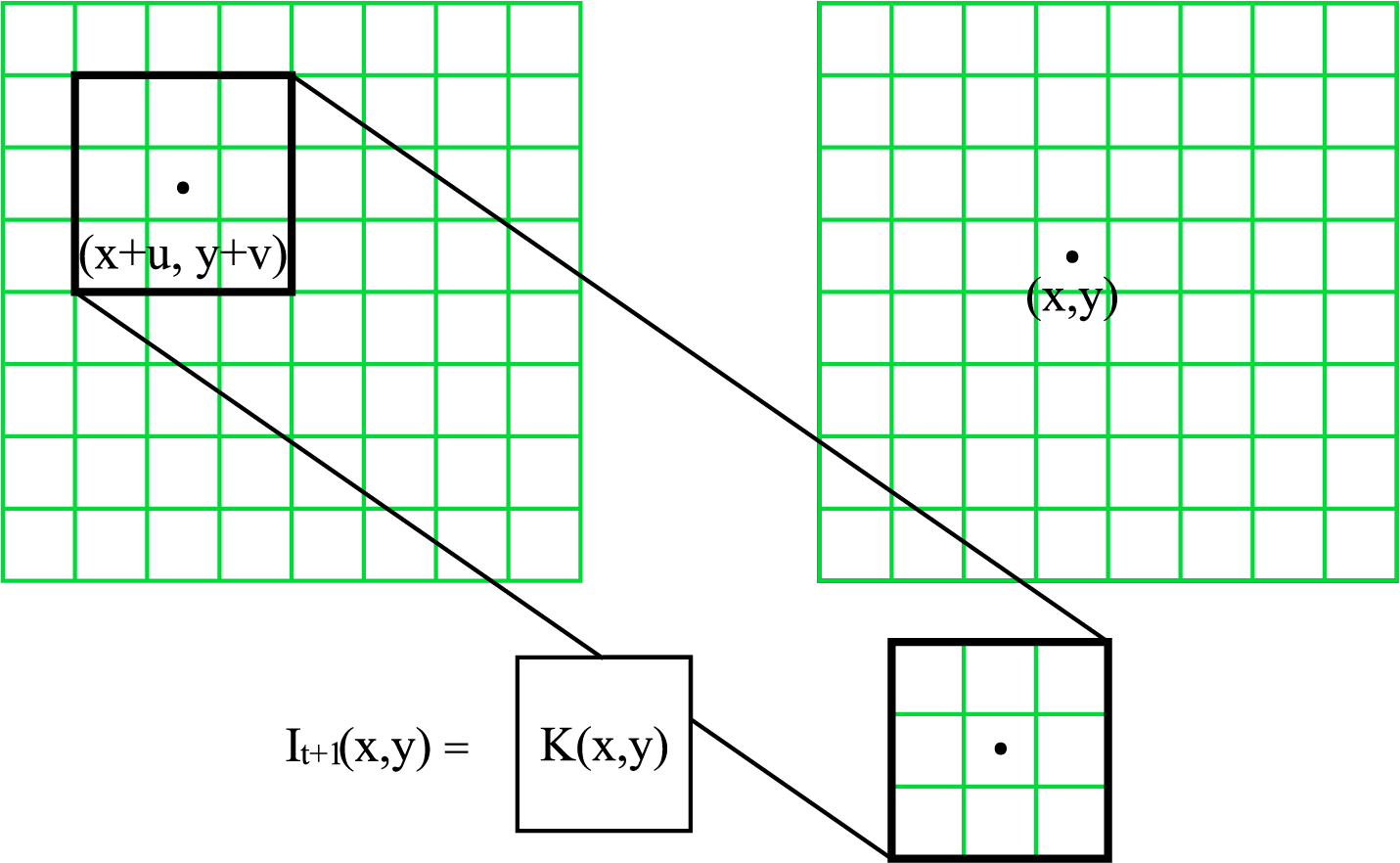}} 
 \end{center}
  \caption{Illustration of Spatial Displaced Convolution.}
 \label{fig:hyper}
\end{figure}

\subsection{Rate Estimation Module}\label{REM}
\indent We model each latent $\hat{y}_i$ as a Laplacian distribution with mean and scale parameters $\mu_i,\sigma_i$  convolved
with a unit uniform distribution. This ensures a good match between encoder and decoder distributions
of both the quantized latents. Based on our solution in \cite{zhou2019end}, both the hyperprior as well
as the causal context of each latent $\hat{y}_i$ are designed to predict the Laplacian parameters. Furthermore, the rate estimation module is a hyperprior network $H$ with parameter $\Theta_{h}$. The predicted Laplacian parameters are functions of learned parameters $\Theta_{h}$:
 \begin{equation}
\begin{aligned} \label{l:py}
   p_{\hat{y}}(\hat{y}|\hat{z},\Theta_{h})=\prod_i(Lap(\mu_i,{\sigma}^2_i)*U(-\frac{1}{2},\frac{1}{2}))(\hat{y}_i),\\
 \end{aligned}
 \end{equation}
where $\mu_i,\sigma=h_d(\hat{z};\Theta_{h})$ is the output of hyperprior network. \\
\indent \textbf{Hyperprior Network} $H$: The subsampled feature $y$ is fed into the hyperprior encoder which summarizes the distribution of standard deviations in  $z=h_e(y)$. $z$ is then quantized $\hat{z}=Q(z)$, compressed and transmitted as side information. The final layer of hyperprior network must have exactly twice as many channels as the bottleneck, so as to predict two values: the mean and scale of a Laplacian distribution for each latent. As to the distribution of $\hat{z}$, we model it as a non-parametric and fully factorized density model because there doesn't exist prior knowledge for $\hat{z}$, , similar to the strategy used in \cite{balle2018variational}:\\
 \begin{equation}
    p_{\hat{z}|\psi}(\hat{z}|\psi)=\prod_i(P_{z_i|\psi_i}(\psi_i)*\mu(-\frac{1}{2},\frac{1}{2}))(\hat{z}_i),
 \end{equation}
where the vector $\psi_i$ represents the parameters of each univariate distribution $P_{z_i|\psi_i}$. \\
\indent Finally, the compression rates are composed of two part: rate $R_y$ of compressed representation $\hat{y}$ and rate $R_z$ of compressed side information $\hat{z}$. These rates are defined as follows:
 \begin{equation}
\begin{aligned}
   &R_y=\sum_i -log_2(p_{\hat{y}}(\hat{y}|\hat{z},\Theta_{h},\Theta_{cm},\Theta_{ep})),\\
   &R_z=\sum_i -log_2(p_{\hat{z_i}|\psi}(\hat{z}|\psi))
 \end{aligned}
 \end{equation}
\subsection{Post Processing}
In order to further improve the performance of our frame, a CNN model is used as our post processing after reconstruction finished. As illustrated in Figure \ref{fig:hyper}, we design the CNN with four layers and the final output with three channels responding to three dimensions of YUV444, the same as the input.
\begin{figure}
   \begin{center}
   \centerline{\includegraphics[width=7cm,height=4.375cm]{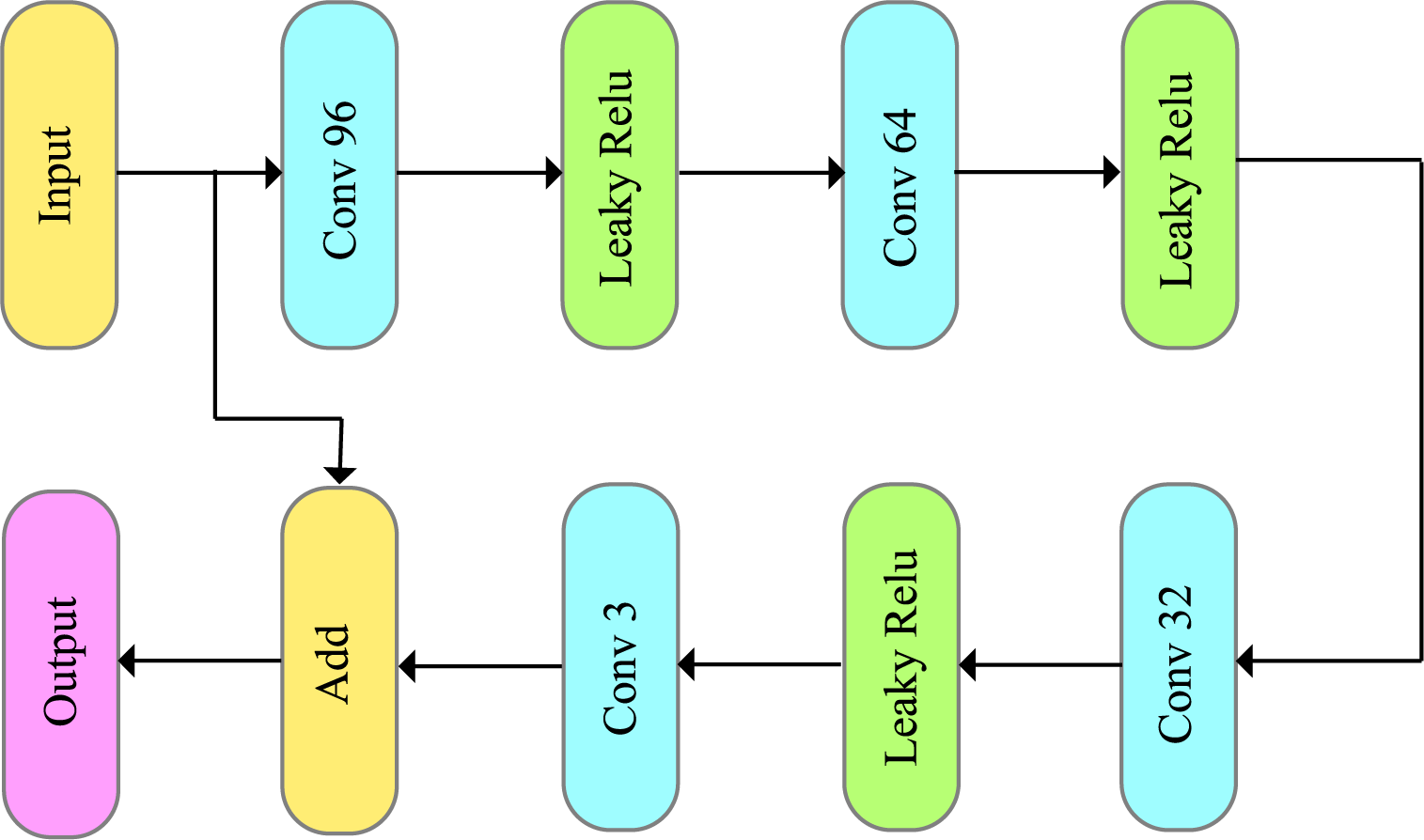}} 
 \end{center}
  \caption{Illustration of Post Processing. Conv n represents the convolution operation with the output channel of n, input is reconstructed pictures.}
 \label{fig:hyper}
\end{figure}
\subsection{Optimized Rate Control}
Rate-Distortion optimization is a common strategy in compression algorithms. The rate-control strategy is similar to our CLIC 2019 solution used \cite{zhou2019end}. Considering the bits constraint, a rate control optimization problem is defined to allocate the bits more effectively for each frame:
\begin{equation}\label{eqop}
   max_{j \in M}~\sum_{i=1}^N MS_j(x_i,\hat{x_i})~~st.~\sum_i~R_j^{i} < R_{max},
\end{equation}
where $MS$ represents the MS-SSIM calculated between original frame $x_i$ and the reconstructed frame $\hat{x}_i$. $M$ is the vector set which contains all possible quality configurations for the set of frames. $N$ is the frame number. $MS_j$ and $R_j$ are the performances and rates under configuration $j$. The best quality configuration is selected for each image via optimizing Eq (\ref{eqop}) in our implementation. The rate control problem is optimized using dynamic programming algorithm.\\

  \begin{table*}
\newcommand{\tabincell}[2]{\begin{tabular}{@{}#1@{}}#2\end{tabular}}
\caption{Evaluation results on CLIC 2020 P-frame validation datasets.}

\begin{center}
\begin{tabular}{|l|l|l|l|l|l|l|}
\hline
   & Methods       &{\bf PSNR} &{\bf MS-SSIM} &{\bf Data Size} &{\bf Decoder Size} &\tabincell{l}{\bf Decoding Time} \\
\hline
&TUCODEC\_SSIM	&37.028 	&0.9969 &	37870015 &	112854058 &	6749 \\
Validation&TucodecVideo	&37.026 	&0.9969 &	37870015 &	84778538 &	N/A \\
&Tucodec	&37.022 	&0.9968 	&37870012&	112847016 &	6954 \\

\hline
\end{tabular}
\end{center}
 \label{t:tvoi}
\end{table*}

\section{Experimental Results}

\indent For training, 463686 image pairs are selected. where each training sample consists of two consecutive frames with the last frame serving as the ground truth. These images are random sampled to 256 $\times$ 256 pixels to train the network. Our team have submitted three solutions: Tucodec, TUCODEC\_SSIM, and TucodecVideo. The results of the validation sets are reported in Table \ref{t:tvoi}. In our implementations, the cluster number is set as 200 for the soft quantization and only one distortion measures perceptional loss are used to train the autoencoder.
\begin{equation}\label{eq1}
  L=\lambda D+ R_y+R_z,
\end{equation}
\indent In TUCODEC\_SSIM the loss $D=1-L_{msssim}$ is defined for the perceptional loss where $L_{msssim}$ is as defined in \cite{wang2003multiscale}. Then the perceptional loss is combined with the same GAN setup defined in \cite{rippel2017real} for network optimization. Then five rates with $\lambda$=20/22/24/26/28 are trained for  
5 rate control. Once the resource allocation is done, MS-SSIM of 0.9969 can be achieved for validation under the constraint of less than 3,900,000,000 bytes with the model size of 112,854,058 bytes. Since there is not enough time, we first submitted a version Tucodec, which is only iteration numbers different from the TUCODEC\_SSIM. Further, we quantize the model with 16 bits, it can compress the model with almost no reduction in accuracy. However, this model have not decoded  successfully on clic server, but it can run perfectly on our local docker. This model achieve MS-SSIM 0.9969 for validation with the model size of 84,778,538 bytes. \\

\section{Conclusion}
In this paper, a novel deep learning based video compression framework which contains a MV-Residual prediction network and a post-processing module is designed for CLIC 2020 challenge. In the MV-Residual prediction network, the motion vectors and residual information are predicted simultaneously. The motion kernels can be learnt by spatially-displaced convolutions to predict pixels in the P-frame by applying the kernels at a displaced locations in the source image. The experiments show that the MV-Residual prediction network can improve the compression performance by modeling the spatial correlation between frames accurately. As shown in the results of the challenges on the validation set, our approaches TUCODEC\_SSIM and Tucodec rank the $1st$ and $2nd$ place in P-frame compression challenge for best MS-SSIM.\\

{ \scriptsize 
\bibliographystyle{ieee}
\bibliography{saliencysegmentation}
}

\end{document}